# High Magnetic Sensitivity at the Coercive Field Induced by Shear Horizontal SAW in Polycrystalline FeGa Films

J.D. Aguilera[1], R. Ranchal[1,2], F. Gálvez[4], J. M. Colino[4], I. Gràcia[3], S. Vallejos[3], A. Hernando[1,5,6,7], P. Marín[1,2], P. de la Presa[1,2,\*], D. Matatagui[1,2,\*]

[1] *Instituto de Magnetismo Aplicado (UCM-ADIF-CSIC), A6 km. 22.500; 28230 Las Rozas, Spain*

[2] *Dpto. de Física de Materiales, Facultad de Ciencias Físicas, Universidad Complutense de Madrid, Plaza de las Ciencias 1, 28040 Madrid, Spain*

[3] *Instituto de Microelectrónica de Barcelona (IMB, CSIC), Campus UAB, Bellaterra, 08193, Spain*

[4] *Departamento de Física Aplicada and Instituto de Nanociencia, Nanotecnología y Materiales Moleculares—INAMOL, Universidad de Castilla-La Mancha, Toledo 45071, Spain*

[5] *Donostia International Physics Center, Donostia 20028, Spain*

[6] *IMDEA Nanociencia, Madrid 28049, Spain*

[7] *Engineering Department, Nebrija University, Madrid, Spain*

*\*Corresponding authors: Patricia de la Presa (pmpresa@ucm.es) and Daniel Matatagui (daniel.m.c@ucm.es)*

## ABSTRACT

A Love wave device was designed to generate surface acoustic waves (SAWs) with strong shear-horizontal polarization, interacting with a polycrystalline $Fe_{72}Ga_{28}$ magnetostrictive layer. The shear strain induced by these waves at a frequency of approximately 160 MHz, coupled with magnetoelastic effects, leads to domain magnetization oscillation, resulting in unique responses that are particularly pronounced near the coercive field. Experimental results reveal that the response of the sensor is highly sensitive to the angle between the applied magnetic field and the wave propagation direction, with profiles that can vary significantly depending on this angle, with some configurations resulting in practically opposite responses. A particularly relevant case arises when the magnetic field is aligned with the Love wave propagation direction. In this case, the sensor response shows mainly a monotonic increase with the magnetic field, except near the coercive field, where a sharp peak emerges and then abruptly collapses, resulting in a magnetic sensitivity of 4.92 Hz/nT. This high sensitivity near the coercive field opens the door to the development of high-performance sensors, simplifying electronics while leveraging the key advantages of SAW technology, including low power consumption, compact size, real-time response, and portability. A theoretical model is also discussed to further understand the underlying phenomena and optimize the design of next-generation devices, which hold significant potential for sensor applications across various fields.

# 1. INTRODUCTION

In the near future, the era of IoT is expected to reach its peak, digitizing our environment and providing us with information through big data while predicting events through AI. Health monitoring,[1] process control in industry,[2] structural monitoring,[3] robot adaptability[4] or environmental research[5] are only a few of the areas in which accurate sensing of physical magnitudes (pressure, temperature, light) or chemical information (gas sensing, biosensing) has become a crucial step for obtaining meaningful results. The research and development of advanced materials, along with a deep understanding of their physical properties, plays a fundamental role in sensor innovation. These materials, combined with carefully designed transducers, determine key factors such as sensitivity, efficiency, and overall performance. However, continued research in the field of materials is essential to achieve a new generation of devices that are significantly more efficient and capable of meeting increasingly demanding technological challenges.

Among the technologies based on the piezoelectric effect, SAWs have played a crucial role in various technological advancements. In these devices, a mechanical wave propagates along the top surface of a substrate. Their ability to precisely manipulate high-frequency signals has made them essential in communication systems,[6] signal processing,[7] and sensing technologies.[8] Different stimuli as temperature changes, mass loading or stress on the substrate can affect wave parameters such as amplitude or phase. Among SAWs we first find Rayleigh waves, the most common type, which involve both vertical and horizontal -longitudinal- displacement in an elliptical motion.[8b] These waves propagate along the surface of a material, generating both compressional and shear forces. However, in addition to Rayleigh waves, we also find SAWs with shear-horizontal (SH) polarization, which have demonstrated advantages in various applications, such as the ability to operate in liquids with low attenuation.[9] To take it a step further, multilayer structures that give rise to guided SH-SAW waves, known as Love waves, can be used to confine energy at the so-called guiding layer surface. By carefully controlling the properties of the guiding layer, most of the energy of the wave remains concentrated near the surface, resulting in a higher amplitude in this region with purely horizontal shear motion. This confinement of energy at the surface allows to efficiently deform any thin layer deposited on the guiding layer. In turn, any variation in this deposited layer will modulate the propagation properties of the wave, making Love wave devices highly effective for sensing applications. Unlike Rayleigh waves, which involve both vertical and longitudinal displacement, Love waves induce displacements that are perpendicular to the propagation direction but entirely confined to the horizontal plane. In summary, the purely shear strain and the confinement of the wave energy of Love waves are the characteristics that make them unique and useful in many applications, and it is especially effective for deforming thin layers.

The exceptionally low bandwidth of the propagated wave in Love wave devices has facilitated the development of highly sensitive sensors, capable of detecting substances at the picogram level through surface functionalization.[10] To achieve selectivity for specific molecular targets, the substrate surface is modified using, for instance, polymer films,[11] antibodies,[12] self-assembled monolayers (SAMs)[13] or nanomaterials like nanoparticles,[14] carbon nanotubes[15] or graphene oxide.[16]

In recent years, Love waves have been utilized to induce deformation in nanostructured layers, enabling the characterization of elastic properties of nanoparticles. This approach also allows for measuring variations in these properties when gases at ppm levels interact with the nanoparticle layer.[17] Considering the magnetoelastic effect, where mechanical stress influences the magnetic properties of a material, it seems ideal to design a Love wave structure with a well-polarized wave. This configuration will be particularly suited for modifying a magnetoelastic layer through shear deformation, with its response expected to depend on the angle of the applied magnetic field. When magnetic sensitivity is required, a natural approach is to use magnetostrictive coatings: these materials exhibit an elongation (shortening) in the direction of magnetization $\vec{M}$ if the magnetostriction constant $\lambda_s$ is positive (negative). Conversely, an applied stress generates a magnetic anisotropy which tends to align $\vec{M}$ with (perpendicular to) the tension. Hence, if magnetization changes during the process, applied stress produces an additional strain. Consequently, the effective Young´s modulus - or shear modulus – varies: this phenomenon is known as $\Delta E$ - or $\Delta G$ - effect. This connection between magnetization and stress has been exploited in the development of magnetic sensors,[17-20] to acoustically stimulate ferromagnetic resonance (FMR),[18] as theoretically suggested by Kittel,[19] and it has also shown to be able to modify the magnetization state of nanostructures[20] or to modulate the necessary field to write or erase magnetic states, described as acoustically assisted magnetic recording.[21] The chosen materials for the coatings have mostly been soft amorphous magnetostrictive alloys, favoring sensitivity to low fields. In magnetic materials many of the most interesting phenomena occur around the coercive field, where magnetization rotation and domain dynamics are most active. This is particularly relevant in magnetoelastic effects, where the interplay between magnetic and mechanical properties is strongly influenced by changes in magnetization near this critical field, enabling advanced sensing and actuation applications. When subject to shear stress, these effects can be further enhanced or modulated, leading to novel functionalities in magnetoelastic devices.

Notably, galfenol alloys are highly relevant in material science due to their magnetostricive properties combined with corrosion resistance, low cost and machinability[22]. In this paper, a Love wave device with a magnetostrictive polycrystalline iron-gallium ($Fe_{72}Ga_{28}$) layer is studied. Previous research has been mainly focused on the alloy $Fe_{82}Ga_{18}$, for which magnetostriction ($\lambda_{100}$) shows a first maximum as a function of gallium content[23]. Alloys with high gallium content have not been extensively investigated, making this study, which explores the alloy $Fe_{72}Ga_{28}$ in the second peak of magnetostriction, particularly significant for exploring their potential applications, even more in SH-SAW devices when this second magnetostriction peak can be caused by an important softening of the tetragonal shear modulus[24]. The combination of Love waves with a polycrystalline magnetoelastic material leads to novel phenomena near the coercive field, where the interplay between elastic and magnetic properties becomes highly nonlinear. In this regime, the coupling between shear-horizontal wave propagation and magnetization dynamics can give rise to unique effects, such as enhanced sensitivity to external stimuli or unconventional magnetoelastic responses. These phenomena open new possibilities for the development of advanced devices, enabling innovative sensing technologies and novel functionalities that go beyond traditional magnetoelastic applications.

## 2. MATERIALS AND METHODS

### 2.1 Love wave device fabrication

The substrate of the sensors consists of a piezoelectric substrate of ST cut crystalline $SiO_2$ with dimensions 4 mm x 9 mm x 0.5 mm. On top of this layer, two symmetric aluminum interdigital transducers (IDTs) with a thickness of 200 nm are patterned by lithography and act as emitter and receiver of the waves, which propagate along the 90ºX crystallographic direction. The period of the IDTs is $\lambda = 28\ \mu m$, the center-to-center separation is $L_{cc} = 150\lambda = 4.2$ mm and the acoustic aperture, which is the length of the strips, is $W = 75\lambda = 2.1$ mm. Four strips per period were patterned, this double (or split) electrode configuration is used to reduce undesired reflections of the generated wave by destructive interference [25]. The chosen substrate, crystal cut and orientation of the IDTs favor the excitation of shear horizontally polarized SAW (SH-SAW) waves.[26] On top of the quartz substrate an amorphous $SiO_2$ layer with a thickness of 3.1 µm is deposited (Figure 1a) by plasma enhanced chemical vapor deposition (PECVD), guiding the SH-waves due to its lower propagation velocity, thus obtaining a Love wave device, with a frequency around 160 MHz, and purely shear strain on the surface. This two-port SAW device features a delay line configuration, enabling measurement of phase and wave losses.

### 2.2 Polycrystalline $Fe_{72}Ga_{28}$ deposition

Once the Love wave device was developed, a thin layer of polycrystalline galfenol ($Fe_{72}Ga_{28}$) was deposited on the space between the two IDTs by sputtering. This precise Gallium-rich alloy was chosen due to its high magnetostriction.[23] For this $Fe_{72}Ga_{28}$ thickness, a magnetostriction constant ($\lambda_s$) of around 73 ppm and an effective magnetoelastic coefficient ($B_{eff}$) of -7.5 MPa is expected [23, 27]. Three different values of the FeGa thickness $t$ were chosen: 50, 100 and 200 nm. A first layer of molybdenum (buffer) was deposited as an interface between $SiO_2$ and galfenol to enhance adhesion, ensure proper crystallization and to avoid oxygen diffusion from $SiO_2$ while a second molybdenum layer (capping) was deposited on top to protect the galfenol from air oxidation (Figure 1a and Table 1). The magnetostrictive $Fe_{72}Ga_{28}$ thin films (50, 100 and 200 nm) were deposited from a target with the same composition by DC magnetron sputtering, using a growth power of 90 W at an Ar pressure of $1\cdot10^{-3}$ mbar. Sputtered Mo buffer and capping layers (5 and 20 nm) were also deposited with a growth power of 90 W and an Ar pressure of $1\cdot10^{-3}$.

| Thickness of the FeGa layer $t$ (nm) | Thickness of the Mo buffer layer (nm) | Thickness of the Mo capping layer (nm) |
|---|---|---|
| 50 | 5 | 20 |
| 100 | 20 | 20 |
| 200 | 5 | 20 |

Table 1: Thickness of the Mo/FeGa/Mo layered structure deposited on the Love wave device.

### 2.3 Structural characterization of the coatings

As the magnetostriction of iron-gallium alloys depends strongly on the crystalline direction, the main structural parameters of the coatings were studied by means of X-ray diffraction (XRD). Measurements were performed in the Bragg-Brentano configuration (θ-2θ) in a D8

Bruker equipment using the Cu Ka wavelength ($\lambda_{CuK\alpha} = 1.54056$ Å). To do this, a 50 nm FeGa film, grown with the same deposition parameters as the coatings presented in this work, but on a glass substrate, was studied. By the application of the Scherrer equation, the crystallite mean size was inferred.

### 2.4 Electrical characterization of the Love wave devices

The scattering parameters of the delay line device based on SH-SAW with magnetostrictive FeGa were studied with a Vector Network Analyzer (Agilent E8326B) as a function of frequency, therefore determining its frequency of operation and associated insertion losses. An oscillator based on a feedback loop circuit was built by connecting both ports and introducing an amplifier that compensates the losses produced in the delay line. This leads to resonant behavior, which was characterized with a spectrum analyzer (Agilent N9010A) by inserting a directional coupler (Mini Circuits ZFDC-20-4) in the loop which allows probing this signal without significant losses. The components of the feedback loop and their arrangement are represented in Figure 1b.

The response of the device consists in resonant frequency ($f_{res}$) shift, which is a result of phase changes of the delay line:[28] following the second Barkhausen criterion, only modes satisfying that the phase shift along the loop is a multiple of 2π can be resonant, therefore, assuming that the rest of the circuit is stable, $f_{res}$ will shift accordingly to these phase changes in order to fulfil this condition. Also, the first Barkhausen criterion must be satisfied - loop gain must equal 1-, forcing $f_{res}$ to remain at the transmission peak. The spectrum analyzer provides around three complete traces per second, which are received by a custom-made LabVIEW program that extracts the resonant frequency. A variable attenuator (Mini-Circuits ZX76-31A-PPS+) is included in the loop circuit (Figure 1b) so the power of the acoustic wave traversing the sample can be varied.

### 2.5 Magneto-optical characterization of the Love wave devices

To better understand the processes involved, the magnetic hysteresis loop $M(H_{app})$ of the FeGa coatings were measured with longitudinal magneto-optical Kerr effect (L-MOKE) magnetometry using a red (632 nm) polarized laser source. Two orientations of the sample respect to the applied field $α$=0°, 45° were explored, where $α$ represents the angle between the SAW wave vector $\vec{k}_{SH-SAW}$ and the applied field $\vec{H}_{app}$ (see Figure 1c). These loops were taken both with Love waves propagating along the sensor and without any excitation.

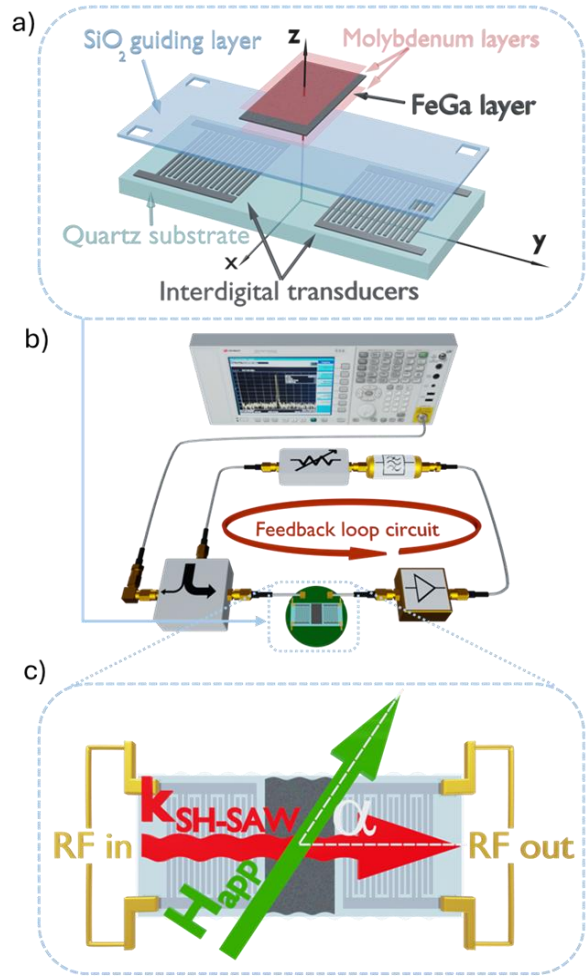

*Figure 1. (a) Schematics of the device and (b) diagram describing the setup of the feedback loop circuit. (c) Zoom on the Love wave device, showing the RF input and output. α is defined as the angle between the in-plane applied field and the acoustic wave vector*

### 2.6 Experimental setup for sensor characterization

To test the response of the device under the application of an in-plane magnetic field with variable intensity $H_{app}$ at different angles $α$, three different systems were used as source:

- **Electromagnet:** the device is introduced between the poles of an electromagnet (Newport Instruments) and exposed to cycles of in-plane applied field while the $f_{res}$ of the circuit is recorded. A custom-made LabVIEW program controls the power supply that excites the electromagnet and stores the data from the spectrum analyzer. This electromagnet was characterized with a magnetometer Magnet Physik FH-55, so the voltage could be converted to field considering the remanence of the ferromagnetic core.
- **Solenoid:** to apply moderate-low fields, the electromagnet is substituted by a cylindric multiturn coil, so no remanence of the core needs to be considered. The excitation, control and calibration systems are the same as those used with the electromagnet.

- **Permanent magnet:** The directional response of the sensor is obtained by exerting the field with a rotating NdFeB magnet (4 x 2 x 1 cm), so the magnitude of the in-plane field stays constant but the angle $\alpha$ that it forms with the wave vector $\vec{k}_{SH-SAW}$ changes. The same magnetometer was used to calibrate the field measured at different distances.

As it usually happens with SAW devices, temperature variations affect the response of the sensor, although using ST (Steady Temperature) cut quartz minimizes the effect. All the measurements shown in this work have been treated to subtract the thermal drift caused mainly by the quartz substrate changing the velocity of SAWs and the wavelength.

# 3 THEORETICAL FRAMEWORK

## 3.1 Quasistatic magnetoelastic model

The proposed 2D model gives some important hints about the *ΔE* effect - change of Young modulus *E* when magnetic field is applied to a magnetostrictive material- when an in-plane magnetic field and shear strain are simultaneously exerted on a magnetostrictive sample.

A monodomain isotropic ferromagnet, with uniform in-plane magnetization $\vec{M_s}$, is subject to an external field $\vec{H}_{app} = H_{app}\vec{u_y}$ ($\alpha = 0°$ for $H_{app} > 0$). Shear stress is also applied on the sample, which interacts with magnetization through magnetostriction, characterized by a magnetostriction constant $\lambda_s > 0$. Shear stress can be decomposed in a compression axis and a tension axis perpendicular to each other, with the same magnitude $\sigma$ and forming an angle $\frac{\pi}{4}$ with the direction of propagation of the shear horizontally polarized wave. Therefore, the induced magnetoelastic anisotropy can be expressed as $k_\sigma = 3\lambda_s\sigma$ in the direction of tension, as two perpendicular opposite anisotropies $k_{\sigma\pm} = \pm\frac{3}{2}\lambda_s\sigma$ are acting at the same time. In absence of shear stress and $H_{app} > 0$, the ideal magnetic material shows a magnetization $\vec{M} = M_s\vec{u_y}$. The elongation in the direction of tension would be $\left(\frac{\Delta l}{l}\right)_{\frac{\pi}{4}} = \frac{3}{2}\lambda_s\cos^2\frac{\pi}{4} - \frac{1}{3}$ if no other than the applied field existed, but once $k_\sigma$ enters into play, the magnetization must rotate an angle θ to minimize the total free energy per unit volume

$$F = F_z + F_{me} = -\mu_0 H_{app} M_s \cos\theta + 3\lambda_s\sigma\sin^2(\theta - \varphi) \qquad [1]$$

In first approximation, assuming that $\theta \ll 1$,

$$\sin\theta = \frac{3\lambda_s\sigma\sin 2\varphi}{\mu_0 M_s H_{app}} \qquad [2]$$

If hysteresis of the ideal sample is considered, for example by introducing an anisotropy field $\vec{H_k} = H_k\vec{u_y}$ in the direction of the applied field, then

$$\sin\theta = \frac{3\lambda_s\sigma\sin 2\varphi}{\mu_0 M_s (H_{app}+H_k)} \qquad [3]$$

Hitherto, the variation in space and time of magnetoelastic anisotropy has not been considered. Tension and compression axes alternate in wave fronts at a distance equal to a half-wavelength. Incorporating this idea into the model, in half the surface of the sample the tension angle is $\varphi$, but in the other half is $\varphi' = \frac{\pi}{2} - \varphi$. Doing the necessary substitutions, it can be seen that expression [2] does not change: $\sin \theta' = \frac{3\lambda_s \sigma \sin 2\varphi}{\mu_0 M_s H_{app}}$. Including in the analysis the varying magnitude of strain does not change the equations, as the elastic and magnetoelastic deformations are linear with $\sigma$.

If the easy axis and the field are not collinear, more complicated expressions must be derived to find $\theta$. Nevertheless, the key difference with the collinear case, important for our analysis of experimental results and given by the Stoner-Wohlfarth model, is the gradual deviation of $\vec{M}$ from field axis as $\vec{H}_{app}$ decreases and reverses. At a certain point, when the energy barrier vanishes, $\vec{M}$ undergoes a reversal settling closer to the direction defined by $\vec{H}_{app}$. However, if the measured hysteresis loops of the samples are well approximated by square ones, the model with collinear anisotropy axis and field can constitute a valid approximation.

In any case,

$$e_m = \left(\frac{\Delta l}{l}\right)_\varphi = \frac{3}{2}\lambda_s[\cos^2(\varphi - \theta) - \cos^2\varphi] \cong \frac{3}{2}\lambda_s \theta \sin 2\varphi \qquad [4]$$

is the magnetoelastic elongation, apart from the purely elastic one, that the wave produces and it is exclusively associated with the rotation of $\vec{M}$.[29] In this approximation, it has been considered $\theta \ll 1$, and it can be seen from Eq. [3] that its dependence with $\sigma$ is linear in that case, as that of the elastic deformation $e_{el} = \frac{\sigma}{E_m}$ where $E_m$ denotes the Young modulus for static magnetization.

The effective Young modulus of the sample $E_{eff}$ must take into account the elastic and the magnetoelastic deformation:

$$\frac{\sigma}{E_{eff}} = \frac{\sigma}{E_m} + e_m \qquad [5]$$

In a good approximation, the frequency of a wave in a solid medium is proportional to $\sqrt{E}$. This is the link between the $\Delta E$ effect and the resonant frequency that we measure.

Four important observations, which will be revisited throughout the text, can be drawn from this simplified model:

(1) Following expressions [2] and [3], $\theta$ and, in consequence, $e_m$, are null for $\varphi = 0, \frac{\pi}{2}$, that is, for $\alpha = 45°, 135°, 225°, 315°$ following the notation used in the experiment. When the tension axis is parallel or perpendicular to the field direction, no deviations of $E_m$ are expected. Conversely, the deviation is maximum for $\alpha = 0°$ and the equivalent directions.
(2) If $H_{app}$ is arbitrarily high, Eq. [2] and [3] show that the orientation of the tension becomes unimportant, as tension anisotropy is unable to deviate $\vec{M}$ from the direction imposed by $\vec{H}_{app}$, and then $E_{eff} \cong E_m$
(3) Although expressions [2] and [3] are not consistent with the assumption $\theta \ll 1$ when $H_{app} \to 0$ or when $H_{app} \to -H_k$ respectively, they suggest that magnetoelastic elongation $e_m$ given by [4] is maximum when reaching those values, as $\vec{M}$ can rotate more freely.

(4) Following the previous idea, an important distinction can be made between the infinitely soft material and the hard one described by Eq. [2] and [3], respectively. In the first case, when magnetization reversal takes place (at $H_{app} = 0$), $e_m$ is the same at $H_{app} = +\Delta H$ and $H_{app} = -\Delta H$, as the physical situation is symmetric. When $H_k$ is significant and $H_{app} \rightarrow -H_k$, $\vec{M}$ deviation is maximum as $\vec{H}_{app}$ and $\vec{H}_k$ are similar in modulus but opposite. Nevertheless, when $H_{app}$ passes that value, $M$ and $H_k$ change sign, and the modulus of the effective field $\vec{H}_{app} + \vec{H}_k$ increases from ~0 to ~$2H_k$, leading to a sudden decrease in $\theta$.

### 3.2 Effects of SH-SAW on FeGa magnetostriction

The most important aspect that the previous model does not take into account is frequency dependence: if shear strain is the effect of a mechanical wave with frequency as high as 160 MHz, a quasistatic model cannot be expected to be totally precise: $\vec{M}$ will not be able to rotate fast enough to reach the energy minimum due to the opposing field generated by currents associated to $\vec{B}$ field changes. Small precession angles, and therefore a linear relation of $\sigma$ and $e_m$, are expected under any circumstance, being this frequency far below the typical FMR frequency of iron-gallium alloys.[30] Anyway, the ideas described remain true and should help describe the phenomenon of SAW modulation.

When dealing with a polycrystalline sample, each grain will have a well-defined anisotropy axis, and an intergrain exchange interaction can be expected. Therefore, the field at which magnetic moments are free to rotate will vary between crystallites, and in fact exchange interaction and magnetostatic interaction will restrict the rotation, so the effect described in observation (3) might be weakened by these constraints. However, the model reveals that for a crystal a key point at which $E_{eff}$ varies significantly is magnetization reversal. If the model is extended to a set of crystals this point will spread over a wider range of applied fields, blurring the sharp step.

Finally, the measured frequency depends on the Young modulus of the magnetostrictive film, but also of the SiO$_2$ substrate (3100 nm) which will not change by itself under the application of a magnetic field.

## 4 RESULTS AND DISCUSSION

### 4.1 Structural characterization

The results of the XRD experiments are shown in Figure 2a. By the observed diffraction peaks, it can be deduced that the texture of the FeGa layer is (110), the same is applicable to the molybdenum buffer and capping layers. By using Scherrer equation, a mean grain size of 7 nm was calculated for the magnetostrictive coating.

### 4.2 Electrical characterization

The characterization of the transmission scattering parameter of the delay line $S_{21}$ as a function of frequency is presented in Figure 2b (red). Just an interval around 160 MHz is shown, as no higher harmonics are observed. The deposition of FeGa produces a shift of the allowed frequencies band towards higher values that increases with the thickness of the film. It also induces additional power losses, reducing the height of the peak. The frequency spectrum in the feedback circuit depicted in Figure 1b is shown in Figure 2b (black): a peak centered in the resonant frequency. The phase selection imposed by Barkhausen criterion gives origin to a narrow spectrum peak, with a *FWHM* (Full Width at Half Maximum) below 50 Hz, that can be tracked in real time as the external magnetic field varies in the scale of seconds or minutes.

### 4.3 Magneto-optical characterization

The results of the magnetic characterization of the magnetostrictive FeGa coatings for angles α=0°,45° between $\vec{H}_{app}$ and $\vec{k}_{SH-SAW}$ are shown in Figure 2c. The *t*=50 nm sample shows a much higher coercive field in both directions. A slight in-plane anisotropy can be observed in the three samples, for instance in the different rates of change of magnetization during the magnetization reversal process.

The magnetization loops showed no changes when acoustic waves power changed, so the magnetism of the FeGa layer is not being significantly affected by the acoustic waves: deviation angles *θ* are small and the interaction is only notably working in the opposite way. More precise measurements would probably show slightly lower coercive fields with SAWs traversing the samples, as the oscillation of local magnetization can help overcome energy barriers, as was found before[21].

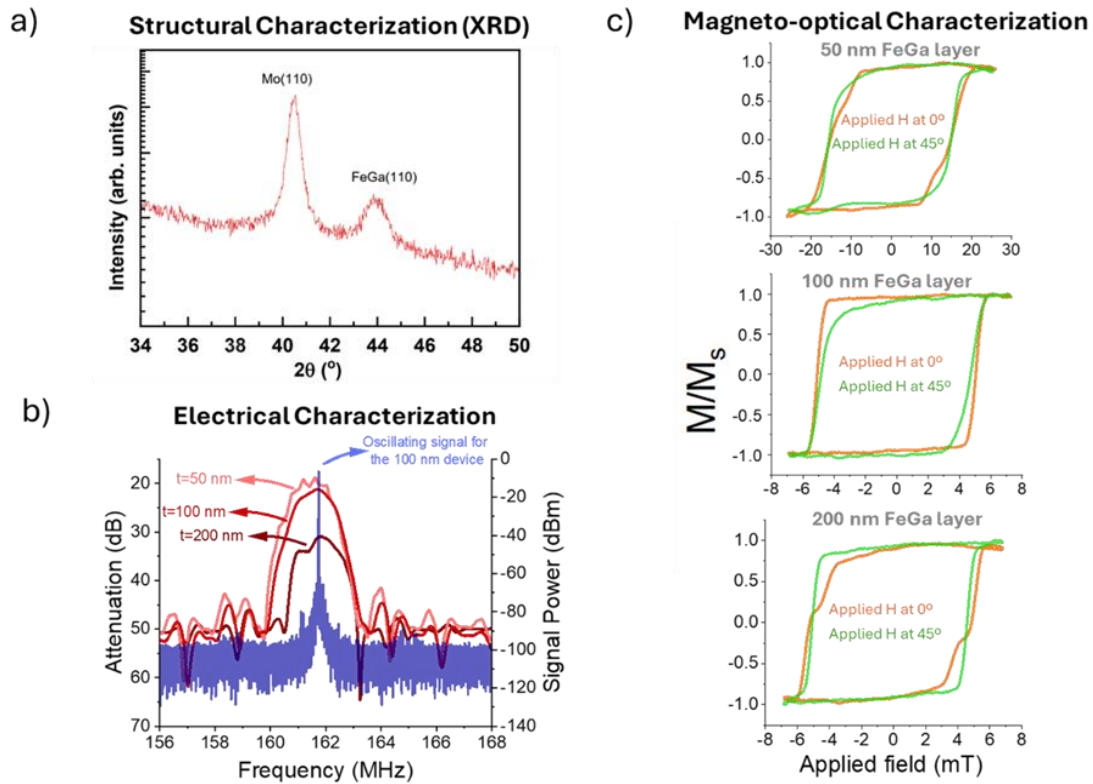

*Figure 2*: *(a) XRD diffraction pattern for a 50 nm-thick FeGa layer. (b) Scattering parameter $S_{21}$ (transmission) for each SAW device (red) and power spectrum of the signal along the feedback loop circuit (t=100 nm) (c) Normalized magnetization loops of the three coatings obtained by L-MOKE at angles α=0°,45°*

## 4.4 Magnetic characterization

The response of the three devices to varying field intensities, starting at $-\mu_0H_{max}\cong-300$ mT and taking the resonant frequency at maximum field as zero reference, can be seen in Figure 3a and 3b for $\alpha=0°$ and $\alpha=45°$ respectively. The insets show the response of the $t=100$ nm device at lower field sweeps, for both increasing and decreasing fields. These two values of $\alpha$ show very different qualitative response at low fields, as published before,[27a] and are selected as representatives of the variety of shapes that the response of the three devices to magnetic fields can assume. As can be observed in both figures, the frequency increases as the field grows above the coercive field, with an apparent saturation-like behavior. The lowest global frequency shift and fastest saturation happens at $\alpha=45°$. This is coherent with observation (1) of section 2.1: this is the angle of magnetization at which shear waves are least effective at displacing magnetization, as anisotropy results to be parallel or perpendicular to it.

The responses of the $t=100$ nm device at angles $\alpha=0°$, $45°$ (Figure 3a and 3b, insets) deserve special attention: upward and downward, respectively, frequency peaks appear around ~5 mT, near the coercive field, according to the magneto-optical characterization (Figure 2c). A sharp change in the frequency shift occurs after reaching the maximum peak, with this drop being more pronounced in the case $\alpha=0°$. With the aim of quantifying this sensitivity – resonant frequency shift with respect to the variation of the magnetic field –, finer measurements on the $t=100$ nm device at $\alpha=0°$ around the coercive field (Figures 3c and 3d) were taken, since the abrupt variation of $f_{res}$ due to the effects of the interaction of the SH-SAW with the magnetostrictive polycrystalline FeGa layer can play a crucial role in the advancement of magnetic sensor technology. For this study, the magnetic field was increased in steps of 260 nT. The maximum frequency shift, 1.2 kHz, was observed just after the peak and the calculated sensitivity at that abrupt peak shown in Figure 3d is 4.92 Hz/nT, in the same order of magnitude of FMR (28 Hz/nT). In order to find the keys of the mechanism of interaction of SH-SAW and the FeGa polycrystalline layer and evaluate the validity of the ideas extracted in section 3, the response -$f_{res}$ shift- and the magnetization $M$ of each device were measured simultaneously under the application of variable magnetic field (Figure S1 of the Supporting Information). The frequency minimum at $\alpha=0°$ (or the maximum at $\alpha=45°$) appears just before the magnetization starts its reversal process, and the frequency shift decreases (increases) during the inversion, finally recovering its smooth response to field. The closeness of the frequency peaks and the coercive field, and the upward peaks at $\alpha=0°$ and downward peaks at $\alpha=45°$ are features shared by the three sensors. The comprehension of how frequency modulation works in polycrystalline deposited layers can lead to the ability of tailoring sensors in which the slope at these peaks is maximized, so exceptionally high sensitivity can be achieved at a particular working point.

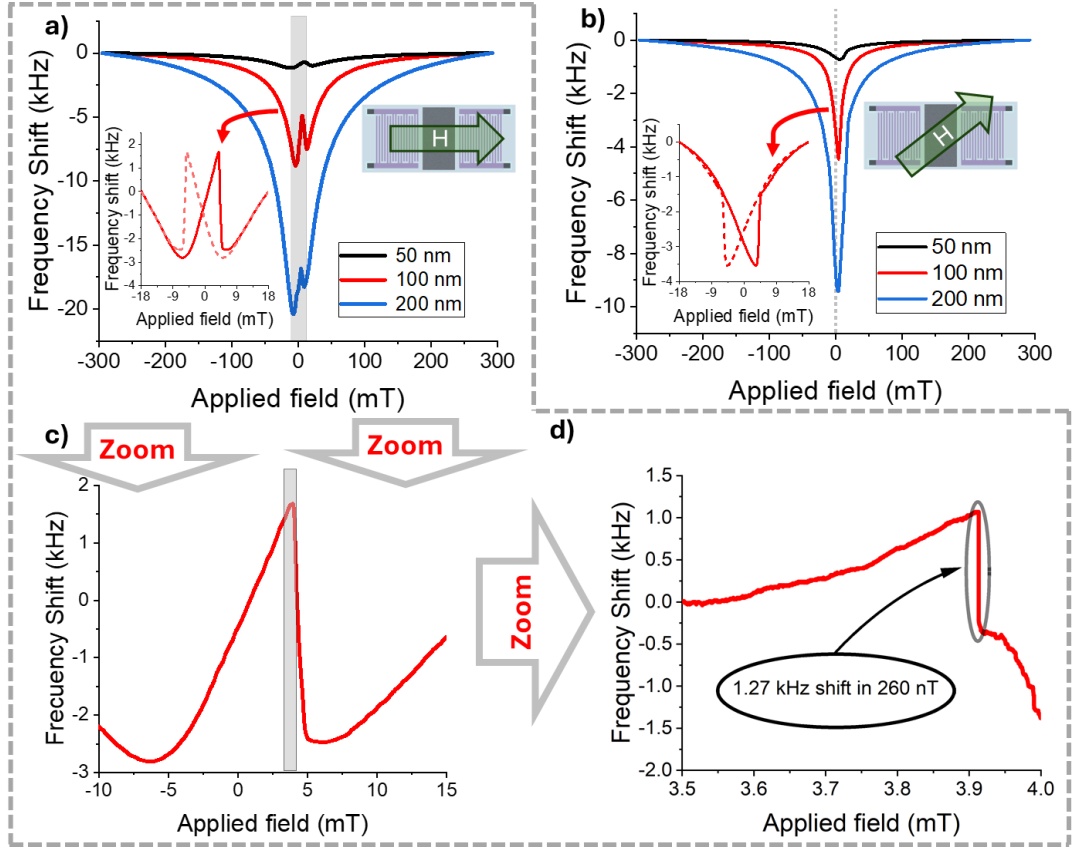

***Figure 3:*** *Response (kHz) of the three (t=50, 100, 200 nm) devices with α=0° (a) and α=45° with field sweeping from -H$_{max}$ to +H$_{max}$ with μ$_0$H$_{max}$=300 mT (b), inset graphs are analogous, with μ$_0$H$_{max}$=18 mT, only the t= 100 nm case is shown. (c) and (d) show different levels of zoom at the response of the t=100 nm device with α=0°.*

The response of the device with the steepest response at the peak, *t*=100 nm, was measured with varying $\alpha$ and constant modulus of the field, for fields ranging from 2.5 mT to 280 mT (Figure 4) to test if the frequency peaks can be attributed to the fast rearrangement of domains suggested in section 3, as observed in the L-MOKE measurements taken simultaneously with the resonant frequency shift. The behavior of frequency shift with varying $\alpha$ is smooth: although there is sensitivity to the direction of the field, no abrupt jumps appear for any field intensity: there is not fast reconfiguration of the magnetic moments, or it is too gradual to be measured. The obtained data have been divided into two graphs to improve visualization: maximum frequency shift at rotation is reached at around 7 mT, decreasing for lower (left graph) or higher (right graph) fields. Another two important hints can be derived from Figure 4:

- In the high field regime, the maximum difference in resonant frequency for a fixed field intensity exerted in any direction becomes lower the higher $\vec{H}_{app}$ becomes, *e.g.* the maximum frequency shift measured at $H_{app}$=288 mT is around 0.5 kHz, and 7 kHz at $H_{app}$=13 mT. The interaction between magnetization $\vec{M}$ and the anisotropy generated by the strain $k_\sigma$ associated to acoustic waves tends to disappear as the applied field increases and locks $\vec{M}$ parallel to $\vec{H}_{app}$ by dipolar interaction, hindering the oscillation. This trend was suggested by the proposed theoretical model (section 3.1, observation 2)

- For any value of the field, at least for fields high enough to neglect the influence of remanence, the maximum resonant frequency is found for $\alpha = 45°, 135°, 225°, 315°$. This is, again, fully coherent with observation 1 of the theoretical model: the deviation of local magnetization must be minimum when it is parallel or perpendicular to the mechanically induced uniaxial anisotropy, which forms 45° with $\vec{k}_{SAW}$. Since the angle $\alpha$, formed by $\vec{k}_{SAW}$ and $\vec{H}_{app}$, and 180°+$\alpha$ are physically equivalent - and symmetry with the measurements from positive to negative field confirm it, in this and in the rest of measurements performed-, only the region 0°≤$\alpha$≤180° is explored. Similarly, due to the symmetry found between angles $\alpha$ and 180°-$\alpha$, only results for angles 0°≤$\alpha$≤90° are shown

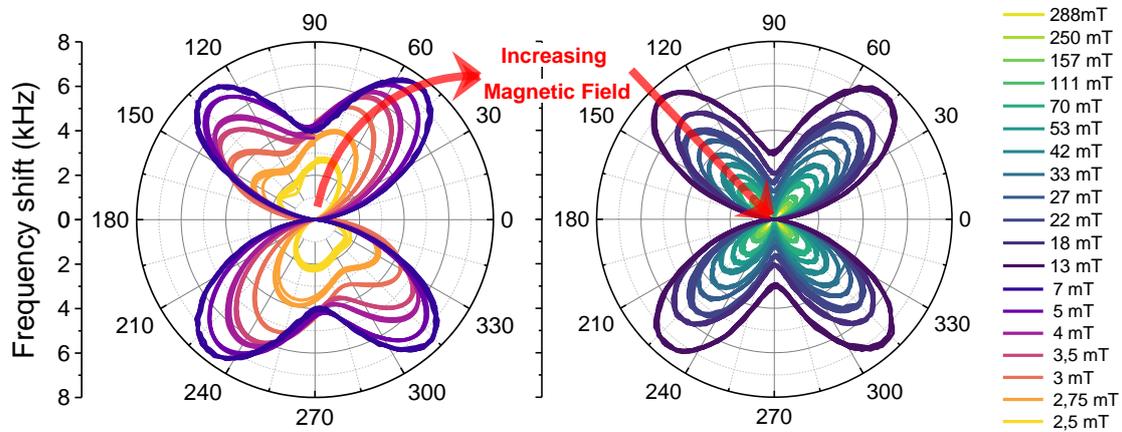

*Figure 4*: Response of the t=100 nm device to rotating fields of magnitudes between 2.5 and 288 mT. The angle of minimum frequency (taken as reference zero) is α=0°.

Figure 5a shows the response of the *t*=100 nm device, at angle *α*=0° to consecutive field cycles, with maximum positive field $\mu_0 H_{max}$=+7 mT and varying in each cycle the maximum negative field -$\mu_0 H_{var}$, converging to zero and starting at -7 mT. In the case of Figure 5b the angle is *α*=45°, and consecutive field cycles, with fixed maximum negative field $\mu_0 H_{max}$=7 mT and varying at each cycle the maximum negative field $\mu_0 H_{var}$, converging to zero and starting at 7 mT. This difference in the cycle measurements was intentionally introduced to easily observe a relationship between the peaks and the hysteresis cycles. In each plot, in red, the measured magnetization loop in that direction is shown. The frequency minimum at *α*=0° (or the maximum at *α*=45°) appears just before the magnetization starts its reversal process, and the frequency shift decreases (increases) during the inversion, finally recovering its smooth response to field. The changes observed when $H_{var}$ crosses the range of maximum slope is particularly interesting: the strong hysteresis observed and the simultaneous vanishing of the peaks suggest a strong link with magnetization reversal: the fast process corresponds to the inversion of magnetic moments respect to the local anisotropy axis, which must be undone when returning to $H_{app}$=-$H_{max}$.

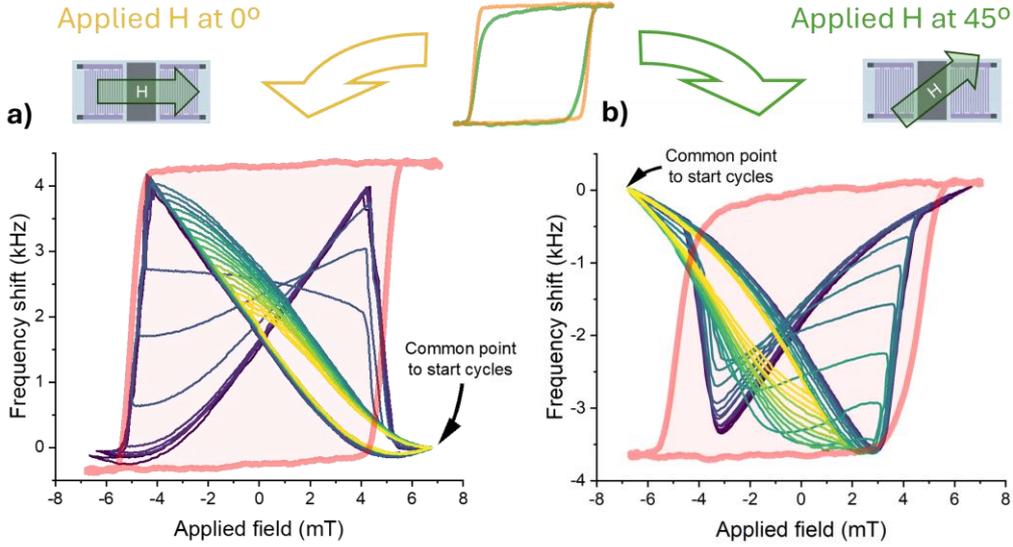

***Figure 5***: *Frequency shift measured from the t=100 nm device when magnetic field cycles are applied with fix maximum field of $\mu_0H_{max}$=+7 mT and decreasing maximum negative field. In red, hysteresis loops measured with L-MOKE at the same angles.*

The ideas exposed in section 3 and the presented data shed light upon the important and intriguing characteristic of upward peaks at *α*=0°,90° and downward peaks at *α*=45° due to the interaction between SH-SAW and FeGa polycrystalline film. The proposed origin is the local disorder of magnetization that appears at low fields: when a high positive field is applied at angle *α*, magnetic moments align with it. As $\vec{H}_{app}$ decreases, local anisotropy dictated by the crystallographic orientation of each grain or group of grains becomes more relevant, and important local disorder appears when $\vec{H}_{app}$ approaches the coercive field. If *α*=0°,90° (*α*=45°), this implies that the orientation of $\vec{M}$ becomes less (more) sensitive to the tension induced by Love waves (see observation 1 in section 3), leading to an increase (decrease) in the effective Young modulus $E_{eff}$ due to the lower (higher) magnetoelastic elongation. In both cases, this effect is maximum just before magnetization in each grain starts switching to the opposite side, close to the coercive field. After that, magnetization will lie much closer to the axis of application of the field (as in the Stoner-Wohlfarth model), and the deviation effect will drastically reduce. These results suggest that peaks appear when domains are not so tightly bound by the external field, particularly when inner effective fields and the external field are of similar magnitude and point in opposite directions, that is, when reaching the coercive field, before magnetization reversal.

Finally, Figure 6 shows the evolution of the response of the *t*=100 nm device to field sweeps from negative values with $\mu_0H_{max}$=18 mT and *α* increasing from 0º (red) to 45º (blue). The reference frequency shift at maximum field for each plot is obtained from the measurements under rotating field (Figure 4). The evolution of the first and second local minima (before and after magnetization reversal) is highlighted by the arrows. The first one happens earlier at *α*=0°, as the effect of disorder (which increases $f_{res}$) opposing to the drop in effective field (which decreases $f_{res}$) is the greatest in this case, and evolves towards fields closer to the coercive field, becoming the downward peak at *α*=45°. The second minimum shows much less variation, as it corresponds to the reorientation of the magnetic moments at the opposite side of the anisotropy axis: disorder is much lower in this case, as the external and internal effective field point in the same direction.

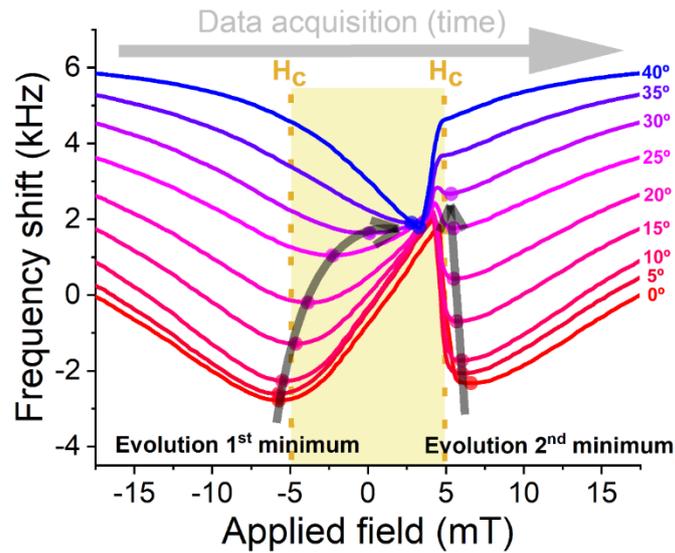

**Figure 6**: Frequency shift of the $t = 100\ nm$ device for magnetic field cycles from $-18\ mT$ to $18\ mT$ and angle $\alpha$ form 0º (red) towards 45º (blue) The reference frequency shift for each angle at maximum field is obtained from rotatory measures (Figure 4b). The evolution of the local minima is highlighted by two arrows.

Finally, based on the results presented in Figure 4, where the results at 0° and 90° are quite different to those at 45°, the measurement range was extended from 0° to 90° in 5° intervals to investigate whether the behavior follows a similar trend (Figure S2 of Supporting Information). At α=0° and α=90°, the peak is consistently upward, with lower intensity at 90°, while it is downward at α=45° and decreases in intensity as we move further from this point for both larger and smaller angles. The specular symmetry of the response with respect to the α=90° direction, observed across the three devices, suggests that the anisotropy induced during growth is unlikely to be the cause of this difference. Although in the formalism presented α=0° and α=90° are considered equivalent, a quantitative difference is observed, which can be explained by two possible factors: (1) the appearance of opposite components of magnetization at the latter angles, creating poles and thus generating a field opposing the magnetization oscillation, or (2) an anisotropy that increases with the magnetoelastic layer thickness, as described in Ref. [27c], making the magnetization reversal process less collective (Figure S3 of Supporting Information) and leading to a "blurring" of the peak, as outlined in section 3.2. The second explanation would account for the increasing difference between α=0° and α=90° observed with thicker layers. This concludes that the optimal angle for operation is α=0°, where the wave propagation and magnetic field are perfectly aligned, maximizing the system's performance.

## 5. CONCLUSION

The use of waves with pure shear horizontal polarization has facilitated the theoretical approach presented in this study, which connects applied tension to changes in the Young modulus of the magnetostrictive layer. This suggests that polycrystalline materials, although magnetically harder than amorphous ones, may exhibit higher sensitivity in specific field ranges (i.e., near the coercive field) due to microscopic anisotropy. The angle of magnetic field regarding the wave propagation affects the peak signal. At $α=0°$, the peak is upward, whereas at $α=45°$, the peak is downward. This behaviour can be explained by the deviation of magnetization in each grain or domain toward the local anisotropy axis, with the ideal case being collinear propagation direction of the wave and the applied field. A more gradual magnetization transition results in a wider peak with a lower slope, as demonstrated by the comparison between hysteresis loops at angles $α=0°$ and $α=45°$. Since the grain size, estimated at 7 nm, is comparable to the expected exchange length, this enables collective magnetization reversal.

The sensor responses were rigorously evaluated under varying magnetic field intensities and angles, with the $t$=100 nm sensor showcasing an exceptionally promising response and a remarkable sensitivity of 4.92 Hz/nT when the magnetic field and wave propagation are aligned. These peaks occur near the coercive field, specifically just before magnetization reversal, when magnetization starts to decrease abruptly. The slope of the response decreases once the inversion is complete, stabilizing the magnetization. Key factors influencing these peaks include film thickness, the angle of the applied magnetic field and the magnetization reversal process.

The findings contribute to the development of sensors using Love waves modulated by magnetostriction, offering ultra-high precision sensing. These next-generation devices hold significant potential for sensor applications across various fields.

## Supporting Information

Supporting Information is available from the Wiley Online Library or from the author

## Acknowledgements

The authors thank the financial supports from Spanish Ministry of Science through the following projects: PDC2022-133039-I00, PID2021-123112OB-C21, PID2021-122980OB-C51 and TED2021-129688B-C21. D. M. acknowledges financial support from the Spanish Ministry of Science and Innovation through the Ramón y Cajal grant RYC2021-031166-I.

## Conflict of Interest

The authors declare no conflict of interest.

## Data Availability Statement

The data that support the findings of this study are available from the corresponding author upon reasonable request.

**Keywords**

Shear Horizontal Surface Acoustic Waves (SH-SAWs), Love Waves, FeGa Magnetostrictive Layer, Magnetoelastic Effects, Sensors, SAW Technology.